\documentclass[journal=jacsat,manuscript=communication]{achemso}
\usepackage{graphicx}
\usepackage{color}
\usepackage{titlecaps}
\usepackage{caption}
\usepackage{amssymb}
\usepackage{amsmath}
\usepackage{gensymb}
\usepackage{booktabs}
\usepackage{multicol}
\usepackage[table]{xcolor}
\usepackage{etoolbox}
\usepackage{natbib}
\makeatletter
\def\acs@contact@details{
E-mail: \acs@email@list
}
\makeatother

\title{Mechanism of Charge Transport in Mixed-Valence 2D Layered Hybrid Bronze Materials}
\setkeys{acs}{
  abbreviations = false,
  articletitle  = true,
  biblabel      = brackets,
  biochem       = false,
  doi           = false,
  etalmode      = truncate,
  keywords      = false,
  maxauthors    = 95,
  super         = true
}
\author{Suchona Akter}
\author{Mohammad R. Momeni}
\email{mmomenitaheri@umkc.edu}
\affiliation{Division of Energy, Matter and Systems, School of Science and Engineering, University of Missouri $-$ Kansas City, Kansas City 64110, MO United States}
\date{}

\begin{document}

\begin{abstract}
Two-dimensional layered bronze (HB) materials are a new class of mixed-valence hybrid organic-inorganic metal oxides that demonstrate great potential as advanced functional materials for next-generation electronics. Recently, new hybrid vanadium bronze materials, (EV)V$_8$O$_{20}$ and (MV)V$_8$O$_{20}$, EV = ethyl viologen and MV = methyl viologen, have been introduced, with EV having $\approx$3 orders of magnitude higher electrical conductivity than the MV system. Given their identical inorganic V--O layers and similar reduction potentials, the observed large difference in electrical conductivities is puzzling. Here, through accurate first-principles calculations coupled with MACE machine learning molecular dynamics (MD) simulations validated by accurate \textit{ab initio} MD simulations, we provide mechanistic molecular-level insights into dominant charge transport and electrical conductivity pathways in these materials. Our detailed structural and electronic properties data identifies factors contributing to this significant difference in electrical conductivities of these materials. Our findings in this work offer clues and provide valuable insights into improving the electrical conductivity of hybrid bronze and similar materials in order to guide the design of next-generation materials with enhanced properties for future electronic and thermoelectric applications.
\end{abstract}


\section{1. INTRODUCTION}
\textbf{1. INTRODUCTION}\newline

The development of efficient and sustainable energy technologies necessitates electrically conductive materials with high charge transport properties. In addition to their high thermodynamic stability and structural integrity, excellent electronic, optical, magnetic, and transport properties of inorganic materials make them highly attractive for practical applications.\cite{zhang20131d} 
Particularly, metal oxides have demonstrated high electrical conductivity, attributed to high charge-carrier mobilities arising from their extended connectivity and high charge-carrier densities.\cite{wan2023controlling} However, fine control over the electrical conductivity of solid-state inorganic materials is a challenge due to the difficulty in directing structure assembly during their energy-intensive synthesis.\cite{wan2023controlling}
On the other hand, organic compounds built from simple molecules are well-known for their flexibility, easy processability, synthesis controllability, and relatively low cost. Incorporation and integration of the two components into a single crystal lattice has led to the creation of hybrid organic-inorganic crystalline materials that possess not only combined or enhanced properties of the individual components but also new phenomena and unprecedented features not possible with either component alone.\cite{kagan1999organic,buhro2003shape,era1994organic,huang2000first}

The structural flexibility of inorganic metal-oxides, combined with their relative insensitivity to stoichiometry and the metallic nature of many of them, has led to the “bronze” nomenclature commonly associated with these materials.\cite{marley2015transformers} The study of inorganic metal-oxide bronzes has spanned decades, revealing a wide range of electronic properties, including variable band gaps, tunable electronic properties, and diverse transport behaviors ranging from semiconducting to metallic and superconducting.\cite{comin2014new,dickens1968tungsten,greenblatt1988molybdenum} These versatile electronic features make these materials highly promising for various energy-related applications such as photovoltaics, electrochromics, batteries, fuel cells, and transistors.\cite{yu2016metal,chernova2009layered,zhang2014overview} As a result, inorganic metal-oxide bronzes serve as excellent parent materials for developing hybrid material platforms.
Among these, binary vanadium oxide V$_{2}$O$_{5}$ is an ideal starting point to understand the structural properties of more complex vanadium oxide bronzes.\cite{galy1992vanadium,enjalbert1986refinement} V$_{2}$O$_{5}$ crystallizes in an orthorhombic layered structure (space group: Pmmn) made up of sheets of alternating VO$_5$ square pyramids, linked by sharing the corners and edges.\cite{marley2015transformers} These layers are bound along the crystallographic c-axis through electrostatic interactions, with interstitial spaces available for cation intercalations. The insertion of cations between these layers is known to modify the framework in response to the size, charge, and polarizability of the cations.\cite{umebayashi2002analysis,yamauchi2005charge,chernova2009layered,banerjee2024synthesis}

The flexibility of the V$_{2}$O$_{5}$ structure is depicted in cross-sectional views of vanadium oxide bronzes, highlighting the adaptability of the open layered framework.\cite{marley2015transformers} The structural versatility of these materials arises from several key features: the accessible V$^{5+}$/V$^{4+}$ redox couple, the stability of various local vanadium coordination environments (such as octahedra, distorted octahedra, square pyramids, and tetrahedra), and the ability to accommodate point defects via crystallographic shear or charge localization/delocalization.\cite{zavalij1999structural,patridge2009synthesis} These factors result in a highly flexible framework, enabling M$_x$V$_2$O$_5$ compounds to stabilize over a broad range of stoichiometries and accommodate cations with varying ionic radii and polarizabilities,\cite{marley2015emptying} while retaining the overall integrity of their frameworks. As mentioned, these cation intercalations significantly influence the electronic and magnetic properties of these materials.

As mentioned, hybrid organic-inorganic materials offer the potential to combine the excellent electronic properties of the extended inorganic framework with the tunability of the organic molecule components.\cite{blancon2020semiconductor} Hybrid bronzes (HBs), a new class of mixed-valence organic-inorganic metal oxides, present a unique opportunity to explore the interplay between molecular structure and electronic properties.\cite{wan2023controlling,n.dayaratne2023hybrid,walte2024mixedmetal} They consist of alternating layers of mixed-valence inorganic metal-oxide layers and ordered arrays of organic molecules. The inorganic layers are structured in an A$_x$MO$_y$ fashion, where A = organic cation and M = Mo, W, V, Nb. HBs have been shown to be synthesizable under milder conditions compared to their inorganic counterparts, often in aqueous solutions at moderate temperatures.\cite{n.dayaratne2023hybrid} They exhibit a metallic luster similar to those found in traditional inorganic bronzes due to the reflection of light by quasi-free electrons.\cite{wan2023controlling} These delocalized electrons are charge-balanced by intercalated cations or oxygen vacancies, resulting in electronic properties ranging from semiconducting to metallic.\cite{wan2023controlling}
The incorporation of organic molecules within HBs introduces several key advantages. Organic molecules are known to act as templating agents, directing the self-assembly of inorganic layers into distinct structural motifs, influencing the degree of electron delocalization, and stabilizing the overall architecture.\cite{n.dayaratne2023hybrid} Moreover, the chemical versatility of organic molecules allows for the introduction of additional functionalities, such as redox or photoactivity.\cite{walte2024mixedmetal}

\begin{figure}[!t]
    \centering
    \includegraphics[width=0.99\linewidth]{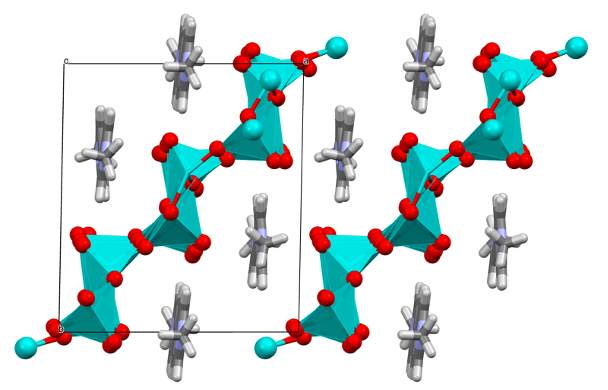}
    \caption{Crystal structure of the (MV)V$_{8}$O$_{20}$ (MV: methyl viologen) hybrid bronze material with anionic vanadium oxide inorganic layers sandwiched between MV organic cations.}
    \label{fig:intro}
\end{figure}

The intricate relationship between molecular structure-directing effects of the templating cations, inorganic topology, degree of reduction, and electronic localization has recently been explored experimentally in a series of eight vanadium-based layered HB materials.\cite{wan2023controlling}
Different diammonium-based templating organic cations were considered by \citeauthor{wan2023controlling}, including two viologen-containing molecules featuring a stoichiometry of (MV)V$_{8}$O$_{20}$ (MV: methyl viologen) and (EV)V$_{8}$O$_{20}$ (EV: ethyl viologen), respectively (Figure \ref{fig:intro}). The vanadium oxide layers within (MV)V$_{8}$O$_{20}$ and (EV)V$_{8}$O$_{20}$ are most similar to those within V$_{2}$O$_{5}$. However, it was demonstrated that the electrical conductivity of (EV)V$_{8}$O$_{20}$ is $\approx$3 orders of magnitude higher than that of the (MV)V$_{8}$O$_{20}$ system,\cite{wan2023controlling} despite both having similar reduction potentials.\cite{michaelis1933viologen} 
The (MV)V$_{8}$O$_{20}$ system, first synthesized in 2001,\cite{bose2002strong} represents a layered organic-inorganic framework with MV molecules confined within the inorganic 2D layers. This system was explored for its potential in novel electrochemical and photochemical energy storage processes and marked the first viologen-containing layered oxide material to be characterized using single-crystal X-ray analysis. In addition, it was deduced that there are strong electrostatic interactions between the aromatic organic molecules and the inorganic layers.\cite{bose2002strong} The intercalation mechanism is facilitated by host-guest charge transfer interactions accompanied by partial reduction of V$^{5+}$ to V$^{4+}$ of the host layer.\cite{zhang1996hydrothermal,bose2002strong} On the other hand, viologen-containing frameworks have also been studied for their excellent thermochromic and photochromic properties.\cite{cadman2021inclusion} Viologen (N,N -disubstituted bipyridinium) species are known to undergo reversible redox reactions that are typically accompanied by distinct color changes.\cite{bird1981electrochemistry} Therefore, they are typically used as redox indicators in biological systems and have more recently gained attention for their potential uses in electrochromic displays and photochemical applications.\cite{madasamy2019viologen} As such, the use of viologen compounds as organic templating cations is expected to play a crucial role in enabling electron delocalization, affecting the overall electronic properties and electrical conductivity of the HB material.\cite{wan2023controlling}  

In this work, the charge transport mechanism in HB materials is studied in detail. Our deduced transport mechanism is then used to provide fundamental molecular-level understanding on the influence of organic molecular templating cations on the topology and flexibility of the inorganic layers as well as overall the electronic properties of these materials. To this end, accurate electronic structure calculations are coupled with the recently introduced general purpose multi atomic cluster expansion (MACE) machine learning models for large-scale MD simulations on nanosecond time scales. Our results show that the closely packed layers in the MV system leads to the opening of the competing out-of-plane charge transport pathway that can interfere and disrupt the main in-plane charge transport pathway. This ultimately leads to a significant reduction of the overall electrical conductivity in MV compared to the EV materials. 
The insights provided by this work will help guide the design and development of next-generation hybrid bronze and similar materials with enhanced electronic properties and functionalities. 
\\\\
\section{2. MODELS AND SIMULATION DETAILS}
\textbf{2. MODELS AND SIMULATION DETAILS}\newline
\subsection{2.1. Electronic Structure Calculations}
\textbf{2.1. Electronic Structure Calculations.}
All first-principles calculations are performed using projector augmented wave (PAW)\cite{kresse1999ultrasoft}, and the Perdew-Burke-Ernzerhof (PBE) density functional \cite{perdew1996generalized} in Vienna Ab initio Simulation Package (VASP 6.4).\cite{kresse1993,kresse1994,kresse1996,kresse1996efficient}  Grimme’s D3 dispersion correction with Becke-Johnson (BJ) damping\cite{johnson2006post} is used in all calculations to account for non-covalent interactions, \cite{grimme2011effect,grimme2010consistent} with periodic boundary conditions applied. Similar to our previous study,\cite{akter2024fine} we conducted a detailed benchmark on minimized cell vectors and atomic positions obtained using different exchange-correlation density functionals, starting from the widely employed PBE functional within the generalized gradient approximation (GGA) formalism; see the Supporting Information (SI) Figure S1. In addition, we tested the performance of the revised version of PBE (RPBE),\cite{hammer1999improved} and PBE for solids (PBEsol),\cite{perdew2008restoring}, combined with a series of dispersion correction schemes in the form of PBE functional with Grimme’s damped D2 and D3 dispersion corrections\cite{grimme2010consistent} as well as the many-body dispersion scheme (PBE-MBD).\cite{ambrosetti2014long,tkatchenko2012accurate} The PBE-D3(BJ) functional was found to closely reproduce the unit cell vectors and atomic positions of the benchmarked {(MV)V$_{8}$O$_{20}$} system compared to the experiment\cite{bose2002strong} with high accuracy. DFT, in its GGA formalism, is known for underestimating band gaps. Hence, the Hubbard U approach (DFT+U), which semi-empirically optimizes the Coulomb potential (U), is adopted to give a better description of the electronic properties. Our extensive benchmarks for the representative {(MV)V$_{8}$O$_{20}$} system show that the DFT+U method, with the Hubbard U, applied to both V and O atoms (3.1 and 8.8, respectively), combined with Grimme’s D3 dispersion correction and BJ damping worked well and closely reproduced that of the experiment (see SI Figure S1).\cite{bose2002strong} This methodology was, therefore, adapted for all studied viologen-containing systems. The plane-wave energy cutoff was set to 500 eV, and all structures were optimized with an electronic energy convergence threshold of 10$^{-4}$ eV, and the magnitude of the largest force acting on the atoms was set to 0.02 eV/\AA. Both atomic positions and unit cell vectors were fully minimized, with the unit cell shape kept fixed (ISIF = 8) at the experimentally determined unit cell. Using fully minimized structures, single-point calculations were then performed with the SCF convergence set to 10$^{-5}$ eV to obtain more accurate electronic properties, including band structures and projected density of states (pDOSs). The smearing method was changed to the tetrahedron, which is shown to provide electronic properties with higher accuracy.\cite{sholl2022density} The $k$-point mesh in the Monkhorst-Pack \cite{monkhorst1976special} scheme was set to 3$\times$3$\times$3 in the geometry optimizations and twice denser for the electronic property calculations. Spin-polarized calculations were performed for all systems.
The BoltzTraP2 code \cite{madsen2018boltztrap2,madsen2006boltztrap} was used for electrical conductivity calculations, which uses a semi-classical method based on smoothed Fourier interpolation of the bands. This method utilizes the constant relaxation time approximation (cRTA),\cite{stanton1987nonequilibrium} assuming that relaxation time $\tau$ is temperature independent and isotropic. This method also assumes the rigid band approach and that crystal momentum is well defined such that the mean free path of charge carriers is larger than the electron de Broglie wavelength.\cite{woods2018assessing}
Calculated thermoelectric properties from BoltzTraP2 were benchmarked at 300 K using a variety of k-point meshes of 6$\times$6$\times$6, 9$\times$9$\times$9, and 12$\times$12$\times$12 for the representative MV system (see SI Figure S9). The k-point mesh of 9$\times$9$\times$9 was selected and used for performing the subsequent transport calculations for the EV systems. The calculated anisotropic electrical conductivities were then partitioned into in-plane (i.e., through bond along the V-O inorganic layers) and out-of-plane (i.e., through space from V-O to the organic molecules) directions for all studied EV and MV systems.

\subsection{2.2. Structure Elucidation from MACE MD Simulations.}
\textbf{2.2. Structure Elucidation from MACE MD Simulations.}
Machine learning interatomic potentials (MLIPs) trained on accurate reference quantum-mechanical calculations are routinely utilized to efficiently predict the energy and forces needed for long-time and large-scale molecular dynamics (MD) simulations.\cite{cheng2024cartesian} In this work, the ASE Python package \cite{ASE_2017} was used for MACE MD simulations in both NVT and NPT ensembles.\cite{batatia2022mace, grunert2025modelling} Starting from the MACE-MP-0,\cite{batatia2023foundation} the second-generation models, including MACE-MP-0b, MACE-MP-0b2, MACE-MP-0b3, MACE-MPA-0, and MACE-OMAT-0 were explored. For these benchmarks, the representative MV system was used with a total simulation time of 1 ns at 300K and 1 atm pressure. The radial distribution functions (RDFs) were then analyzed by comparing to the reference AIMD data (SI Figure S5); more details on reference AIMD simulations are provided in the next section. 
MACE MD simulations in the NVT ensemble employed a time step of 1 fs and a Nose-Hoover chain thermostat of chain length 3 and a time constant of 100 fs. Among different tested MACE models, the MACE-MP-0-Small model demonstrated better performance in terms of accurately capturing the electrostatic interactions between the organic and inorganic components as deduced from our RDF analyses (see SI Figure S5) and was therefore chosen for all considered systems. Subsequently, using the chosen MACE model, MACE MD simulations in the NPT ensemble were performed on the larger 2$\times$2$\times$2 supercells allowing the volume to change. The Berendsen thermostat and barostat (inhomogeneous) as in Ref. \citenum{juraskova_modelling_2025} was used for these simulations. The first 500 ps of all MACE MD NPT simulations was discarded as equilibration with the second portion used for statistical analyses.

\subsection{2.3. AIMD Reference Data for MACE Validations.}
\textbf{2.3. AIMD Reference Data for MACE Validations.}
Periodic AIMD simulations were carried out at the PBE-D3-BJ level in CP2K\cite{cp2k} employing a mixed Gaussian and plane-wave basis set as implemented in the quickstep module.\cite{vandevondele2005quickstep} Norm-conserving Goedecker-Teter-Hutter (GTH) pseudopotentials were used for core electrons, \cite{goedecker1996separable,hartwigsen1998relativistic,krack2000all} while the valence electron wavefunction was expanded in a double-$\zeta$ with
polarization basis set (DZVP-MOLOPT-SR-GTH).\cite{vandevondele2007gaussian} Considering the rather large size of the studied HB materials, the plane-wave cutoff energy was reduced to 300 Ry.
All AIMD simulations were performed in the canonical (NVT) ensemble for all EV2, EV3, EV4, and MV systems at 298.15 K with a time step of 1.0 fs. The temperature was kept constant using the canonical sampling through velocity rescaling (CSVR) thermostat\cite{bussi2007canonical} with a time constant of 200 fs.\cite{mones2015adaptive} The 0K geometry optimized structures were adopted as the initial configuration for all AIMD simulations. All AIMD simulations were performed for a total of 20 ps, where the first 15 ps was discarded as equilibration, and the last 5 ps was used for statistical analysis. Energy and temperature convergence plots from AIMD simulations are given in the SI Figure S3.
\\\\
\section{3. RESULTS AND DISCUSSION}
\textbf{3. RESULTS AND DISCUSSION}\newline

\subsection{3.1. Structural Properties and Flexibility of HBs from MACE MD Simulations.}
\begin{figure}[!t]
    \centering
    \includegraphics[width=0.85\linewidth]{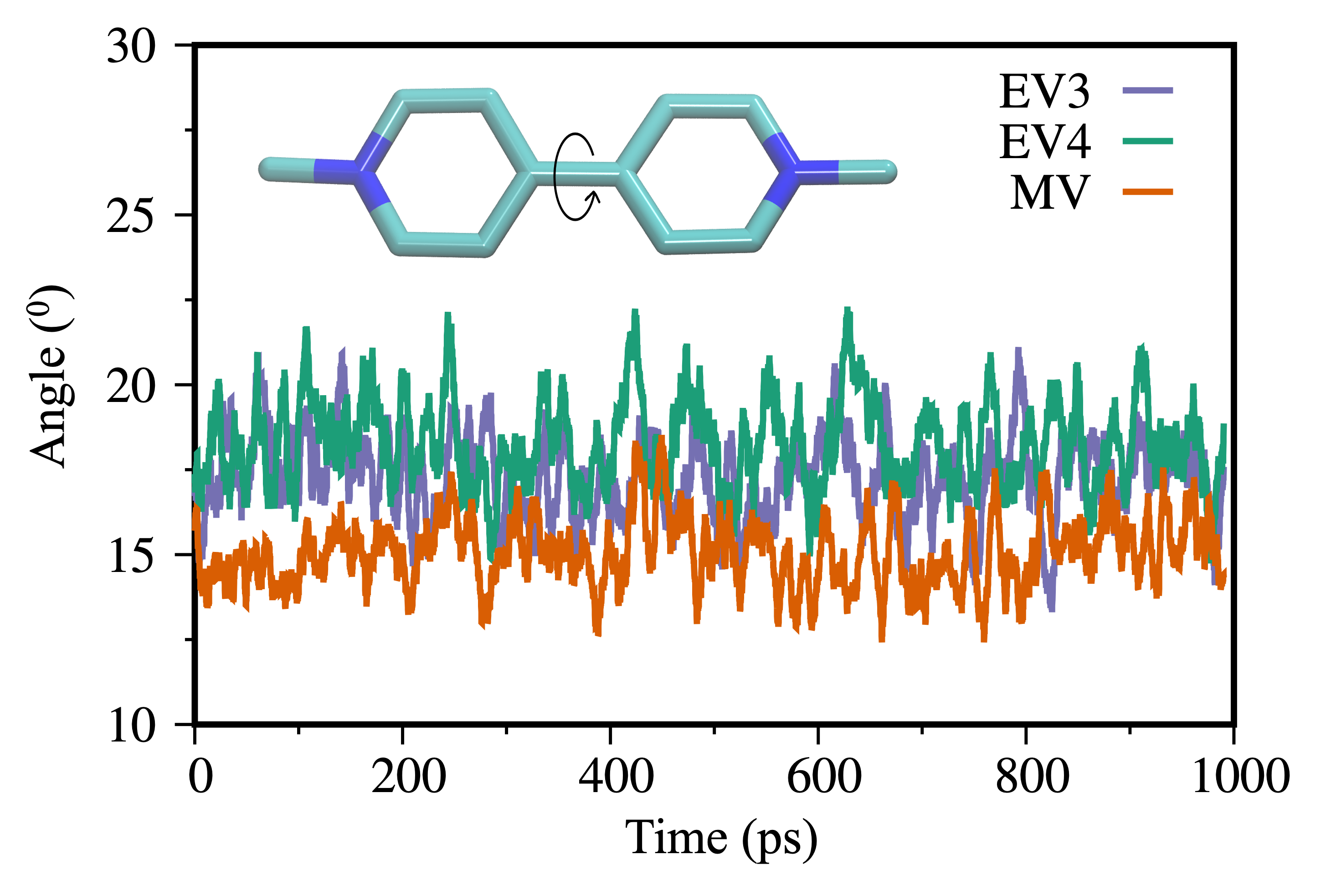}
    \caption{The MACE MD analysis of the dihedral angles formed between the aromatic rings in EV3, EV4, and MV systems.}
    \label{fig:dihedral_angle}
\end{figure}

\textbf{3.1. Structural Properties and Flexibility of HBs from MACE MD Simulations.}
Our 1ns MACE MD simulated data are used to probe the degree of flexibility or rigidity of the organic molecules in different EV and MV systems using the dihedral angles formed between the two aromatic rings (Figure \ref{fig:dihedral_angle}). MV molecules were found to form an average of 15.1$^\circ$ angle, whereas the corresponding angles are 17.3$^\circ$ and 18.3$^\circ$ for EV molecules in EV3 and EV4 materials, respectively. These angles confirm the co-planarity of the MV molecules stemming from the less steric hindrance of the methyl groups vs. those of ethyl in EV systems. This could potentially lead to a greater inorganic layer separation in the EV systems than the MV, weakening the electrostatic interactions between the cationic organic and anionic inorganic layers.

\begin{figure*}[!t]
    \centering
    \includegraphics[width=0.7\linewidth]{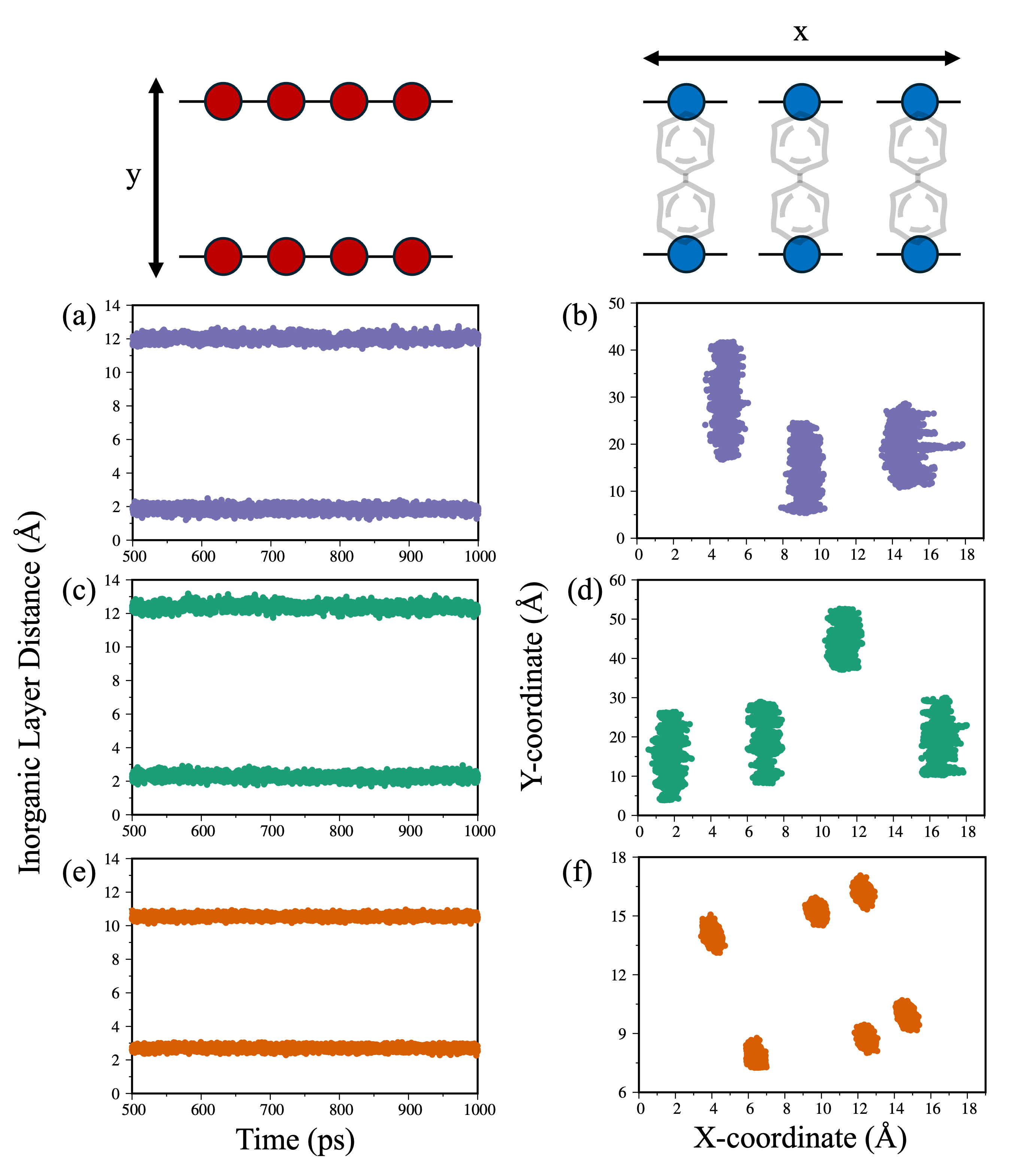}
    \caption{Calculated flexibility of the inorganic (left) and organic (right) layers in EV3 (a-b), EV4 (c-d), and MV (e-f) systems as obtained from MACE MD simulations. The positions of the V and N atoms, schematically represented by orange and blue spheres, are used for these analyses.}
    \label{fig:flexibility}
\end{figure*}

The inorganic layers in both the EV and MV systems exhibit a certain degree of flexibility (Figure \ref{fig:flexibility}). This flexibility likely arises from the readily accessible V$^{5+}$/V$^{4+}$ redox couples and the stability of various local vanadium coordination environments, including octahedral, distorted octahedral, square pyramidal, and tetrahedral configurations.\cite{marley2015transformers} As shown in Figure \ref{fig:flexibility} (a), in EV-containing systems, the inorganic inter-layer distances are greater than those in MV. The presence of bulky ethyl groups in the EV3 and EV4 systems leads to a greater separation of the V–O inorganic layers, which could result in anisotropic charge transport in these systems. This suggests that the likelihood of electron transport along the out-of-plane direction is lower in the EV systems than in the MV. In contrast, the closely packed layers in MV could lead to a competing charge transport via the out-of-plane channel, potentially disrupting the in-plane charge transport pathway and ultimately reducing the overall electrical conductivity in this material. This disruptive effect should be notably less significant in EV systems. In the next sections, we will discuss how these structural data dictate different electronic properties in these materials.

To quantitatively probe the strength of the host-guest interactions in the considered HB materials, the binding energies of all considered EV and MV systems were calculated as:\cite{schwalbe-koda2021benchmarkinga}
\begin{equation}\label{eqn:binding}
    E_{bind} = E_{tot} - E_{inorg} - n \times E_{org}
\end{equation}
where $E_{tot}$ is the DFT calculated total energy of the system, with $E_{inorg}$ and $E_{org}$ being the energy of the inorganic and organic components and $n$ the number of organic molecules in the material.
The calculated binding energies are summarized in Table \ref{tab:binding}, with more negative values indicating stronger host-guest interactions.

\begin{table}[h]
    \centering
    \begin{tabular}{cccc}
        \hline
        Systems & $E_{bind}$ (eV/molecule) & m$^*_e$ & m$^*_h$\\
        \hline
        EV2 & -6.61 & 11.25 & 32.71\\
        EV3 & -7.35 & 6.18 & 61.64\\
        EV4 & -7.19 & 7.66 & 10.12\\
        MV& -7.48 & 33.30 & 10.71\\
        \hline
    \end{tabular}
    \caption{Calculated binding energies per organic molecules ($E_{bind}$, eV/molecule) along with effective masses for electrons and holes (m$^*_e$ and m$^*_h$) for all considered systems.}
    \label{tab:binding}
\end{table}

As can be seen from Table \ref{tab:binding}, the calculated $E_{bind}$ values first increase as $n$ increases going from EV2 to EV3 but then slightly decreases in EV4. Considering all systems, our calculations show that MV has the highest binding energy per organic molecule of all. Given the inorganic layer is stoichiometrically identical in all EV and MV systems, this observation supports the conclusion that layers in MV are more closely packed with stronger host-guest interactions, whereas in EV, the layers are more separated due to the greater steric hindrance of the larger ethyl groups.

\begin{figure*}[!ht]
    \centering
    \includegraphics[width=0.85\linewidth]{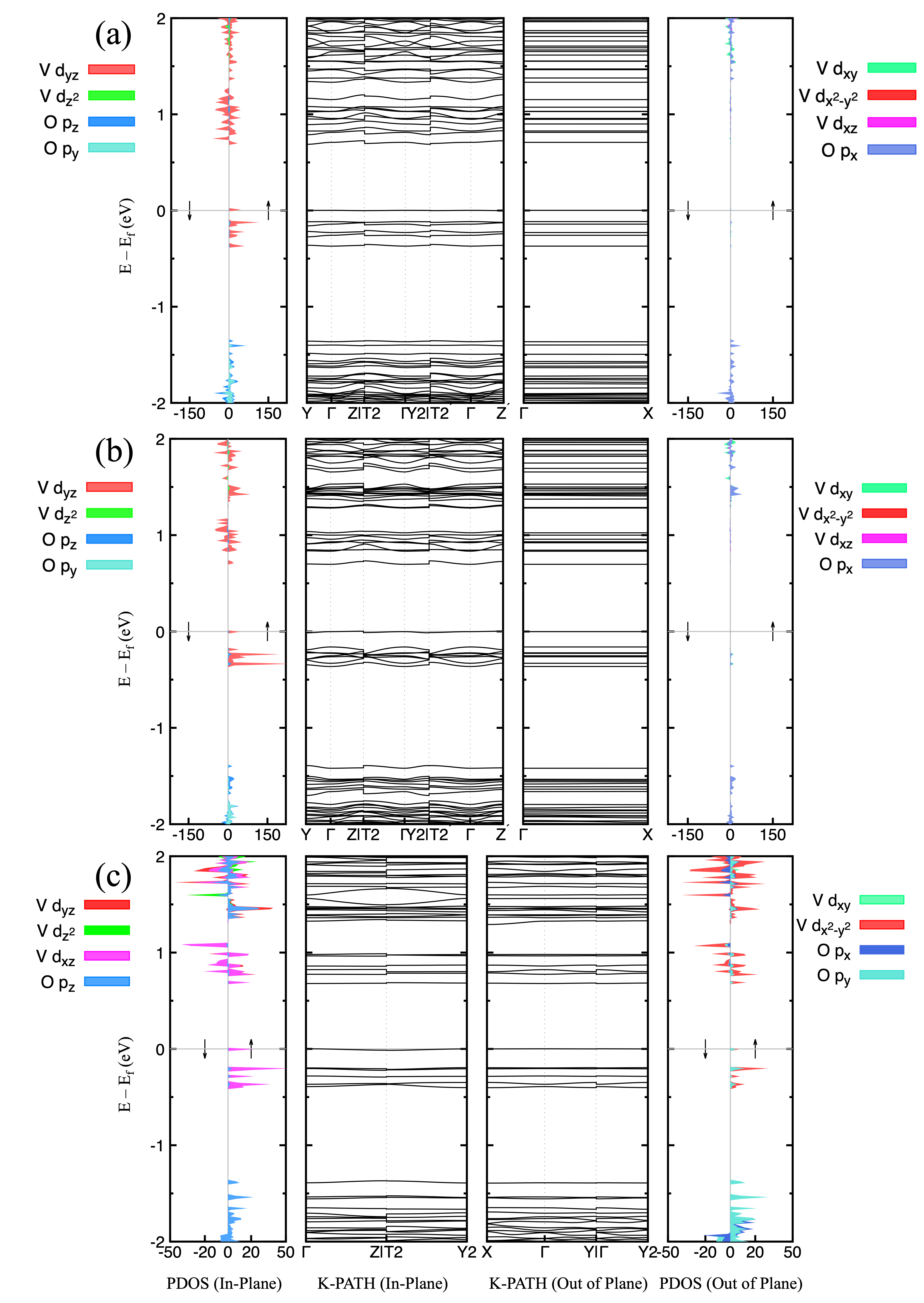}
    \caption{Calculated in-plane vs. out-of-plane band structures and projected density of states for (a) EV3, (b) EV4, and (c) MV systems. The k-point paths are given in the SI Figure S8.}
    \label{fig:band-dos}
\end{figure*}

The calculated radial distribution functions (RDFs) as obtained from our MACE MD simulations were used to further analyze the different host-guest interactions in all considered systems (see SI Figures S4 and S5). Calculated RDFs of all considered pairs were found to be very similar for both the EV3 and EV4 systems. The calculated V-O RDFs show two peaks around 1.6 and 2.0 \text{\AA}. Comparing MV to EV, the calculated RDFs between organic and inorganic components were found to be shorter in MV than in EV systems (see SI Figure S4). This analysis again indicates that the inorganic layers in MV are more compact and have stronger electrostatic interactions with the organic components than in EV systems. As mentioned, in the EV-containing systems, the ethyl group in the bipyridine ring induces greater steric hindrance, causing the layers to be more separated compared to MV. This strongly corroborates that electrons should be more delocalized in EV-containing systems, which is beneficial for enhancing electrical conductivity.

\subsection{3.2. Electronic Properties.}
\textbf{3.2. Electronic Properties.}
To better understand the mechanism behind the significantly enhanced electrical conductivity in the EV systems compared to that of MV, a comparative study of their electronic band structures, pDOSs, and electron and hole effective masses was conducted. The band gap was calculated to be 0.7 eV for all considered HB systems showing semiconducting behavior for these materials. In band structures, the dispersion of the bands is defined as the energy difference between the lowest and highest points of a band at two distinct k-points, which reflects the ease of charge carrier transport across the material. A greater dispersion indicates more mobile charge carriers, stronger overlaps, and extended interactions along that crystallographic direction. The effective mass of a charge carrier in a band can be obtained by calculating the second derivative of the band near high-symmetry points in the k-space,\cite{mancuso2020electronic} according to the following equation:
\begin{equation}
m^* = \hbar^2 \left( \frac{\mathrm{d}^2 \varepsilon(k)}{\mathrm{d}k^2} \right)^{-1}
\end{equation}
where $\epsilon(k)$ are the eigenvalues at the band edges around the conduction band minimum (CBM) and valence band maximum (VBM), with $k$ being the wave vector. A smaller effective mass illustrates a higher mobility of the electron and hole charge carriers. The calculated Band structure and density of states are plotted in Figure \ref{fig:band-dos} with effective masses given in Table \ref{tab:binding}. 

Two in-plane and out-of-plane charge transport pathways are identified for all HBs considered in this work. In MV, the in-plane path was identified as $\Gamma$-Z-T$_2$-Y$_2$, while the out-of-plane direction was found to correspond to X-$\Gamma$-Y-$\Gamma$-Y$_2$ k-point path (see SI Figure S8). For the EV systems, the Y-$\Gamma$-Z-T$_2$-Y$_2$-T$_2$-$\Gamma$-Z path was identified as the in-plane path, while $\Gamma$-X was found to correspond to the out-of-plane direction.
 \begin{figure*}[ht]
    \centering
\includegraphics[width=0.99\linewidth]{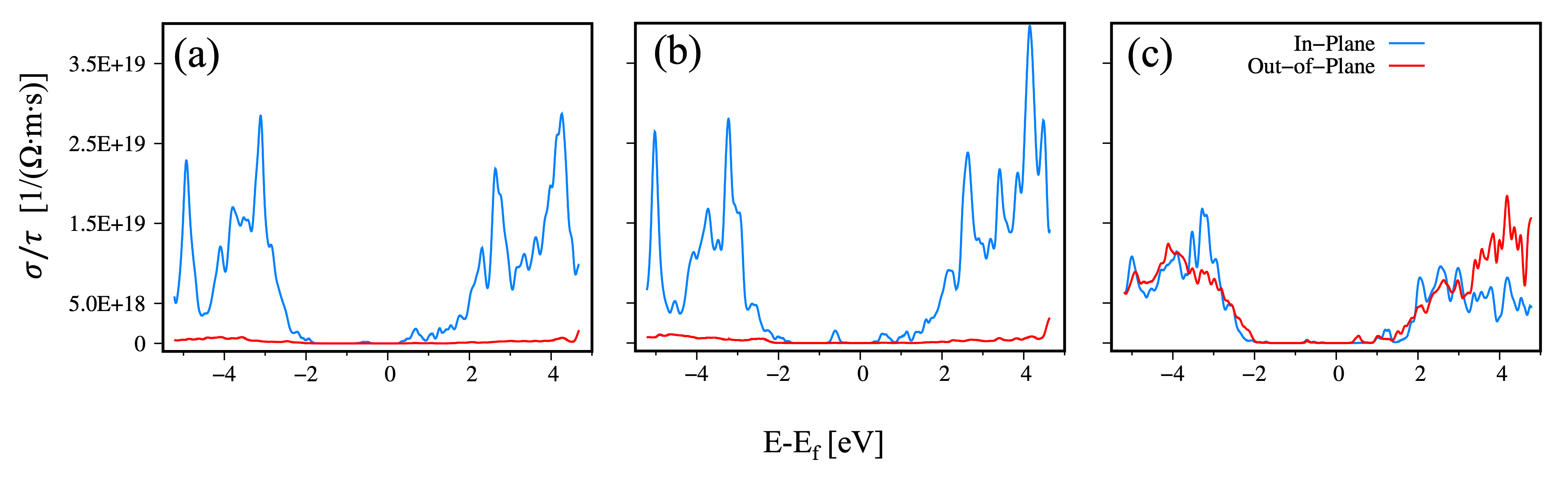}
    \caption{Calculated anisotropic electrical conductivity in (a) EV3, (b) EV4, and (c) MV systems.}
    \label{fig:conductivity}
\end{figure*}
The MV system is found to exhibit relatively flat VBM and CBM, suggesting relatively large effective masses and small electron and hole mobilities. Meanwhile, in the EV systems, the VBM and CBM are more dispersed than the MV, which qualitatively agrees with the higher electrical conductivity of the EV systems compared to that of the MV. A detailed examination of the in-plane and out-of-plane band structures shows a distinct trend in the EV system, where the electron density is significantly higher in the in-plane direction compared to that of the out-of-plane (Figure \ref{fig:band-dos}). Notably, the out-of-plane bands exhibit a flat dispersion, which indicates a high effective mass and restricted electron/hole transport.
Conversely, in the MV system, the out-of-plane charge density is comparatively higher than in the EV system. This increased out-of-plane charge density introduces scattering effects that disrupt electron transport in the in-plane direction, ultimately leading to a reduced electrical conductivity in this material. In contrast, the EV systems experience minimal out-of-plane disruptions, allowing for directional charge carrier transport along the lateral pathway and overall higher electrical conductivity.

\subsection{3.3. Anisotropic Electrical Conductivity}
\textbf{3.3. Anisotropic Electrical Conductivity.}
To assess the origin of the anisotropic electrical conductivity in HBs, electrical conductivity along two major in-plane and out-of-plane pathways are studied for all considered EV and MV systems (Figure \ref{fig:conductivity}). The calculated in-plane electrical conductivity for the EV systems is significantly higher than that of the out-of-plane. In contrast, the in-plane and out-of-plane electrical conductivities are comparable in MV. It is known that two transport mechanisms, including band transport and hopping can contribute to the overall charge transport and electrical conductivity of different materials. For example, in vanadium oxides, ligand-to-metal charge transfer can occur from O 2p orbitals to V 3d orbitals within isolated molecular units whereas hopping of the relatively localized electrons occurs through space between the neighboring redox-active sites.
In the latter conduction mechanism, charge transport occurs through hopping via non-covalent interactions (e.g., $\pi–\pi$ stacking)\cite{xie2019diverse}. This mode of charge conduction has already been observed in similar 2D layered electrically conductive metal-organic frameworks (MOFs) where charge carriers are known to easily hop between redox-active fragments along the inter-layer directions with distances of around 3.5~\AA.\cite{pccp_23_3135,aplm_9_051109,acsami_13_25270,jcp_156_044109,acsami_15_9494} In the case of 2D layered HB materials, aromatic benzene rings in the organic molecules can transiently accept electrons from the inorganic layers, which may allow for electrons to transport in the direction perpendicular to the inorganic layers (i.e., the out-of-plane pathway). The overall contribution of this through space charge transport pathway is expected to strongly depend on the interlayer distances between the inorganic layers as dictated by the templating cations.

As mentioned, our data, as shown in Figure \ref{fig:conductivity}, clearly indicate that the contribution of in-plane is significantly higher than that of out-of-plane for the EV systems. This agrees with our calculated electron and hole effective masses for the EV4 system, which is the lowest among all, allowing for efficient electron transport without significant disruption, as the layers are well-separated due to the larger steric hindrance.
In contrast, in MV, which demonstrates the largest electron effective mass of 33.3 among all (Table \ref{tab:binding}), the in-plane and out-of-plane contributions to the electrical conductivity are calculated to be comparable (Figure \ref{fig:conductivity}). This agrees with our dihedral angle analysis depicted in Figure \ref{fig:dihedral_angle}, which revealed that MV remains more coplanar compared to EV, allowing for more compact inorganic layers. This compactness causes the inorganic layers to get closer (see Figure \ref{fig:flexibility}), which creates competition between out-of-plane and in-plane charge transport pathways. This competition ultimately leads to an overall reduced charge mobility and lower electrical conductivity in MV compared to EV systems.
\\\\
\section{4. CONCLUSIONS AND FUTURE WORK}
\textbf{4. CONCLUSIONS AND FUTURE WORK}\newline

The tunability of HBs highlights the great potential of these materials as ideal candidates for the next generation of robust hybrid organic/inorganic electrically conductive materials. In this study, we demonstrated that electroactive organic molecular guests, such as methyl and ethyl viologen, significantly affect the structure and, in turn, electrical conductivity properties of the parent vanadium oxide materials in agreement with the available experiment. The detailed analyses presented in this work reveal the intricate charge transport mechanisms in play for methyl vs. ethyl viologen organic templating cations, providing explanations on how small structural changes can induce around three-orders-of-magnitude difference in electrical conductivity. Guided by the insights provided in this work, our future studies will focus on developing a comprehensive database of HB materials with improved electronic properties spanning a range of semiconductor to metallic materials for different applications.

\begin{suppinfo}
Benchmarks of different exchange-correlation functionals with and without dispersion and Hubbard U corrections, details of structure assignments for the EV system including PXRD patterns and calculated RMSEs, E and T convergence plots for AIMD simulations, E, T, and V convergence plots for MACE MD simulations, calculated RDFs from MD simulations using different MACE models compared to that of the reference AIMD, calculated flexibility plots using MACE MD simulations vs. AIMD, and k-point convergence plots for anisotropic electrical conductivity calculations.
\end{suppinfo}

\begin{acknowledgement}
M.R.M thanks Farnaz A. Shakib for many stimulating discussions. Simulations presented in this work used resources from Bridges-2 at Pittsburgh Supercomputing Center through allocations PHY230099 and CHE240033P from the Advanced Cyberinfrastructure Coordination Ecosystem: Services \& Support (ACCESS) program,\cite{access} which is supported by National Science Foundation grants \#2138259, \#2138286, \#2138307, \#2137603, and \#2138296. The authors also acknowledge the HPC center at UMKC for providing the use of computing resources and support.

\end{acknowledgement}

\Addlcwords{is with of in the an a iv v as for on and by to at}

\bibliography{bib,SI/bib}

\begin{tocentry}
\includegraphics[width=0.99\linewidth]{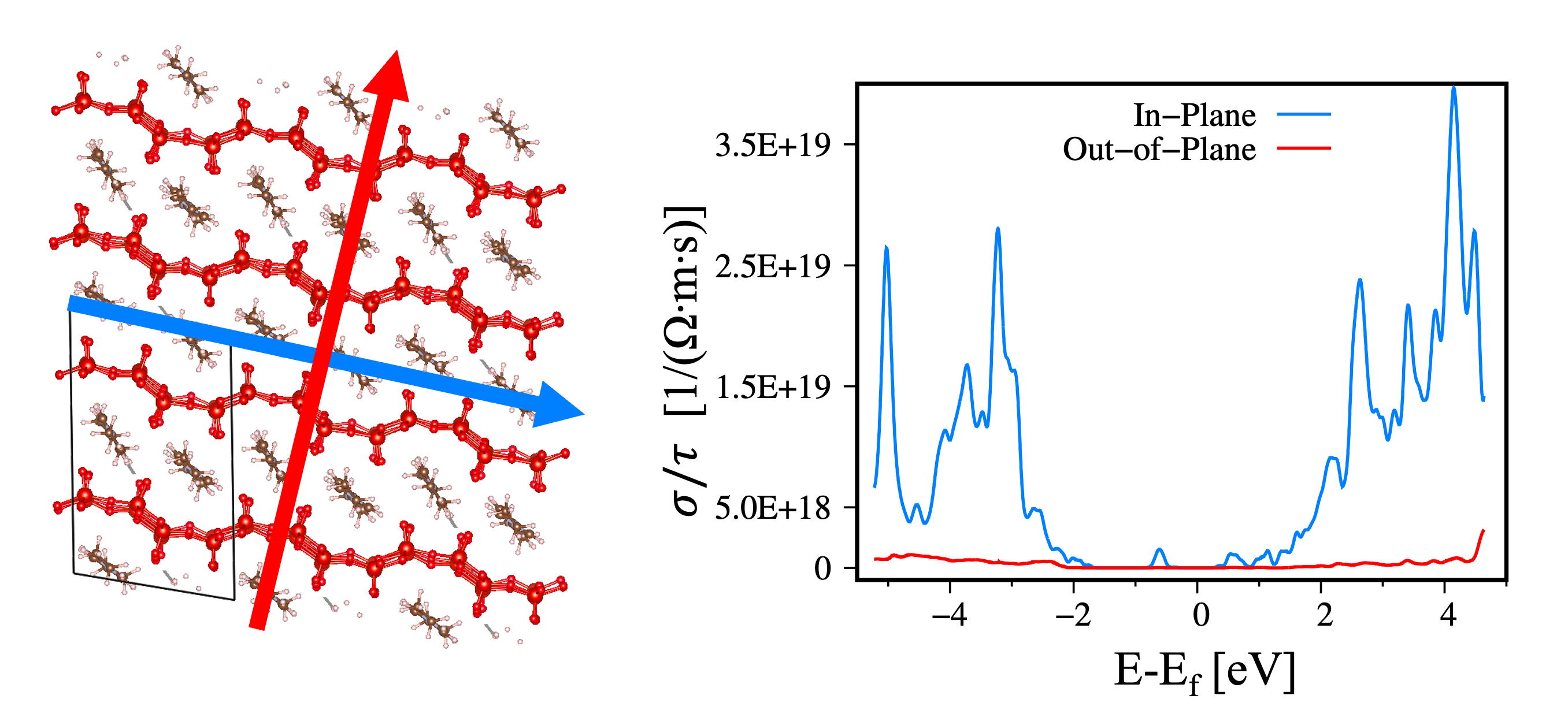}
TOC entry
\end{tocentry}

\end{document}


\newpage
\tableofcontents
\newpage

\begin{figure*}[!h]
    \centering
    \includegraphics[width=0.99\linewidth]{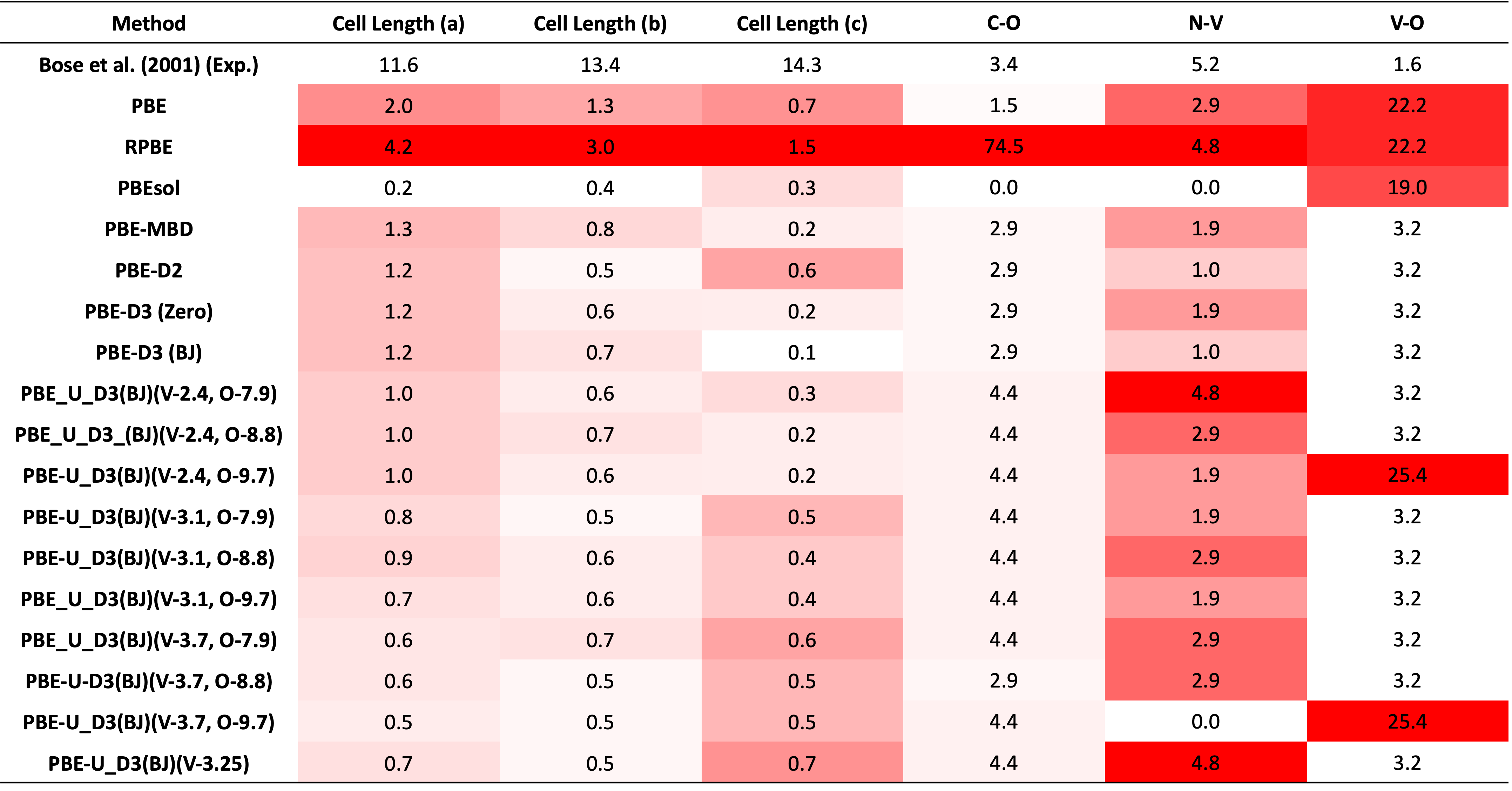}
    \caption{Calculated \% error in unit cell lengths and distances for the (MV)V$_{8}$O$_{20}$ system using different exchange-correlation functionals with and without dispersion and Hubbard U corrections compared to the experiment.\cite{bose2002strong}}
    \label{fig:benchmark_mv}
\addcontentsline{toc}{subsection}{Figure \ref{fig:benchmark_mv}. Benchmark of different exchange-correlation functionals for MV.}
\end{figure*}

\clearpage
\section{1. Assigning the Structure of (EV)V$_8$O$_{20}$}\label{sec-1.1}
\addcontentsline{toc}{section}{1. Assigning the Structure of (EV)V$_8$O$_{20}$}
In the reported (EV)V$_{8}$O$_{20}$ crystal structure, the EV molecules could not be resolved.\cite{wan2023controlling} Therefore, the first step in our theoretical analyses is to determine the precise location of these EV molecules within the crystal structure. Experimentally, a diffuse chain of electron density was observed, suggesting that the EV cations are likely located within this channel of electron density. To account for this, half-occupancy carbon atoms were positioned at the regions of the highest electron density, which exhibited a ribbon-like pattern. The chemical composition also indicated the presence of nitrogen, which is not explicitly included in the model. Alternative methods, such as SQUEEZE, were not used because the observed ribbon of electron density provided clear evidence for the presence of a flat, elongated molecule that is not confined to a single position within the lattice.

Here, we compared the experimental PXRD patterns with our simulated curves to verify the correct number of EV molecules in this system. As such, different structures with varying numbers of EV molecules were created to match the PXRD patterns for (EV)V$_8$O$_{20}$. We designed unit cells containing only a single EV, followed by structures with 2, 3, and 4 EV molecules. Due to the lattice symmetry, a complete EV molecule was not present in the unit cell as derived from the experimental CIF file. By extending the unit cell along the b-axis by a factor of four, we were able to accommodate the larger number of EV molecules considered in this work.

After constructing the extended unit cell, the stability of the EV1-containing system was evaluated using AIMD simulations. Between 4 and 6 ps, the layered architecture of this material was found to unravel, which led us to discard the EV1 structure. Subsequent calculations were then focused on unit cells containing 2, 3, and 4 EV molecules. Figure S2 presents the calculated PXRD patterns for all EV systems and their corresponding RMSE values compared to that of the experiment. The EV2 system was discarded due to its comparatively higher RMSE value and unfavorable binding energy (see the main text for more details).
Our detailed analyses considering PXRD patterns, binding energies and dynamical data as obtained from extensive MACE MD NPT simulations reveal that both EV3 and EV4 are plausible candidates for the experimentally reported EV system. We note that the difference between EV3 and EV4 is simply in the degree of which the V sites are reduced and therefore depending on the experimental conditions we anticipate both systems to be viable candidates.

\clearpage
\begin{figure*}[!h]
    \centering
    \includegraphics[width=0.99\linewidth]{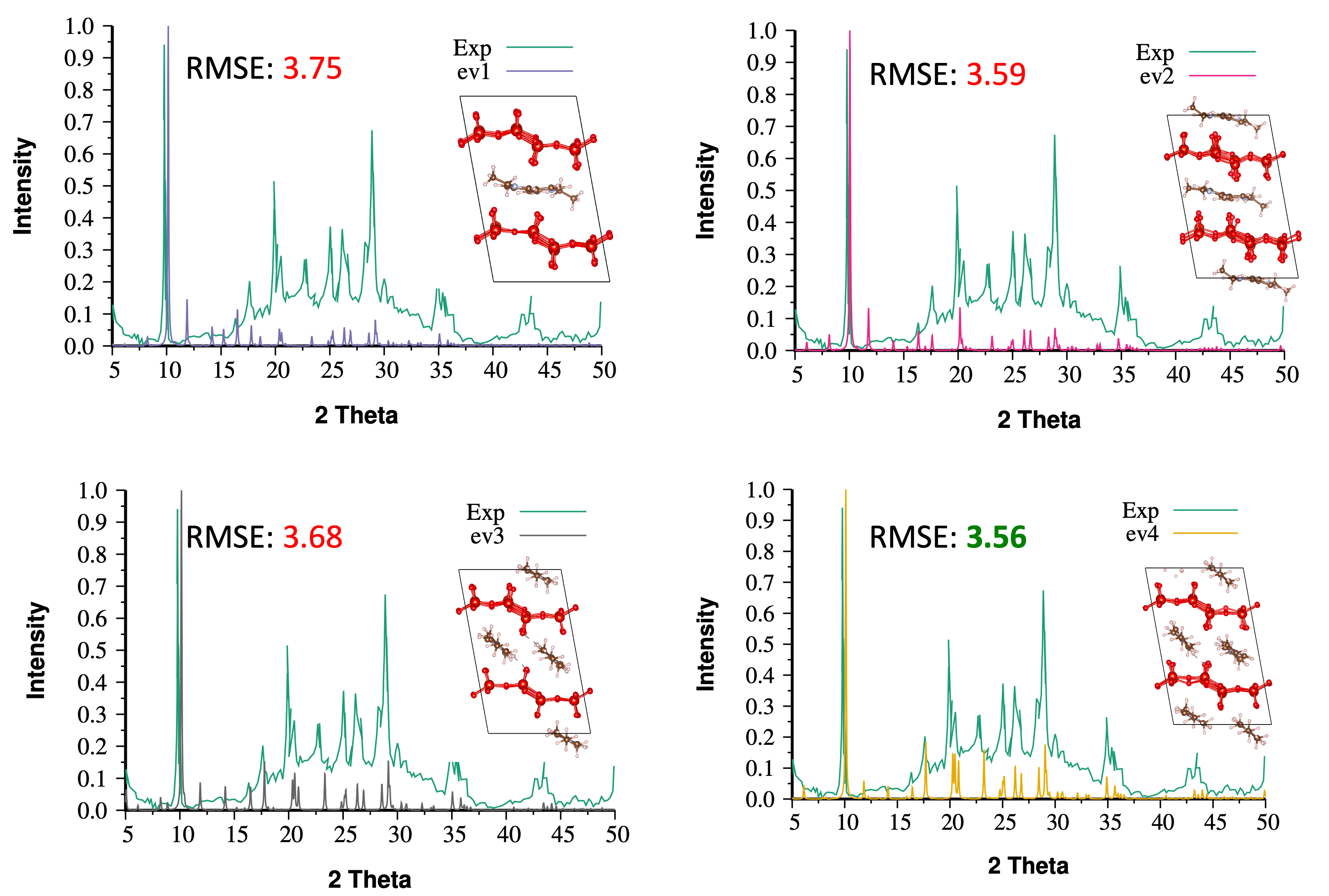}
    \caption{PXRD comparisons among HBs with different numbers of EV molecules as obtained from DFT optimized structures compared to that of the experiment.\cite{wan2023controlling}}
\addcontentsline{toc}{subsection}{Figure \ref{fig:pxrd}. PXRD comparisons among HBs with different numbers of EV molecules.}
    \label{fig:pxrd}
\end{figure*}

\clearpage
\begin{figure*}[!h]
    \centering
    \includegraphics[width=0.99\linewidth]{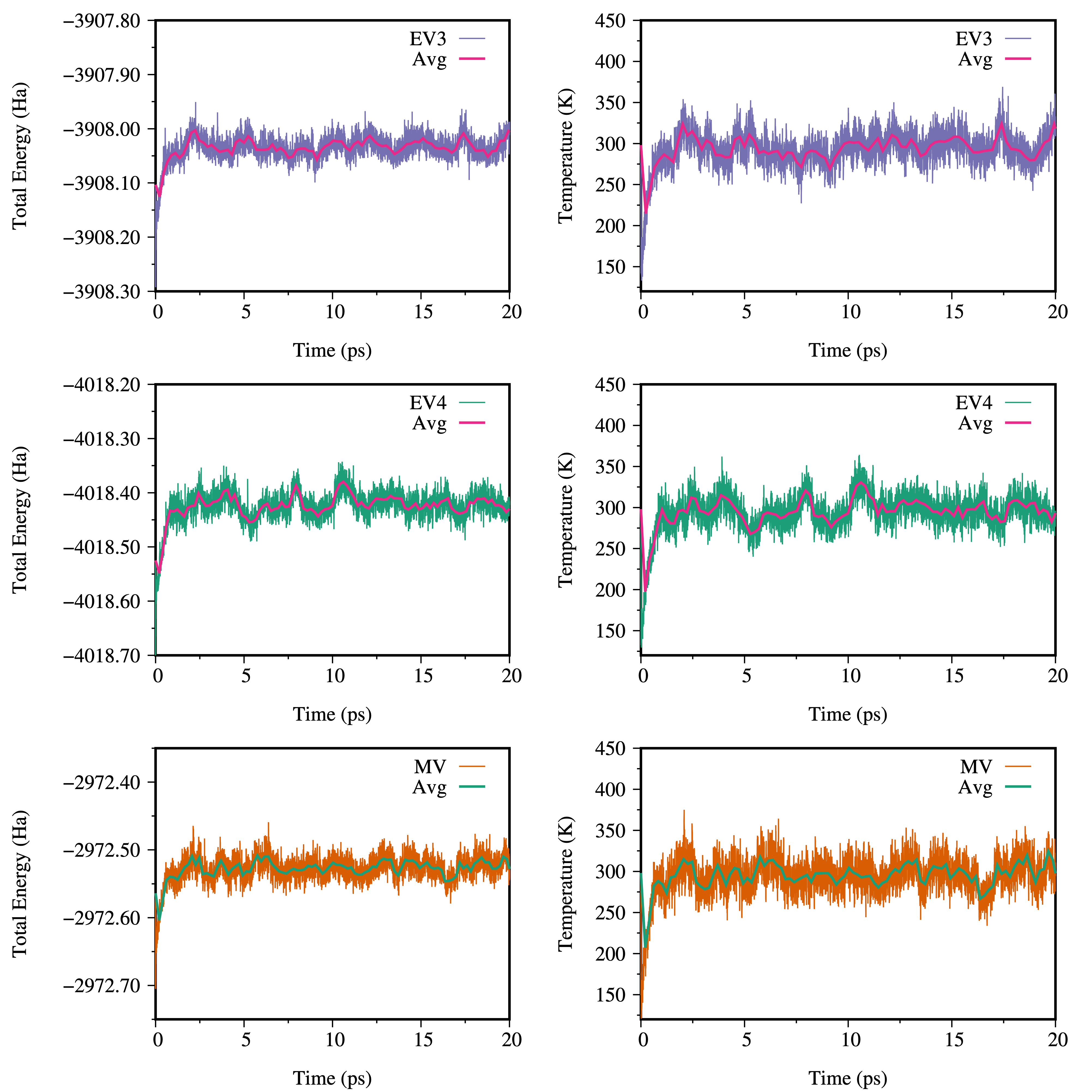}
    \caption{Convergence of the energy (in Hartree) and temperature (in kelvin) for the EV3, EV4, and MV systems at 298.15 K in the NVT ensemble as obtained from reference AIMD simulations. The moving 50 points average is shown in red.}
\addcontentsline{toc}{subsection}{Figure \ref{fig:aimd-conv}. AIMD convergence plots.}
    \label{fig:aimd-conv}
\end{figure*}

\clearpage
\begin{figure*}[!h]
    \centering
    \includegraphics[width=0.99\linewidth]{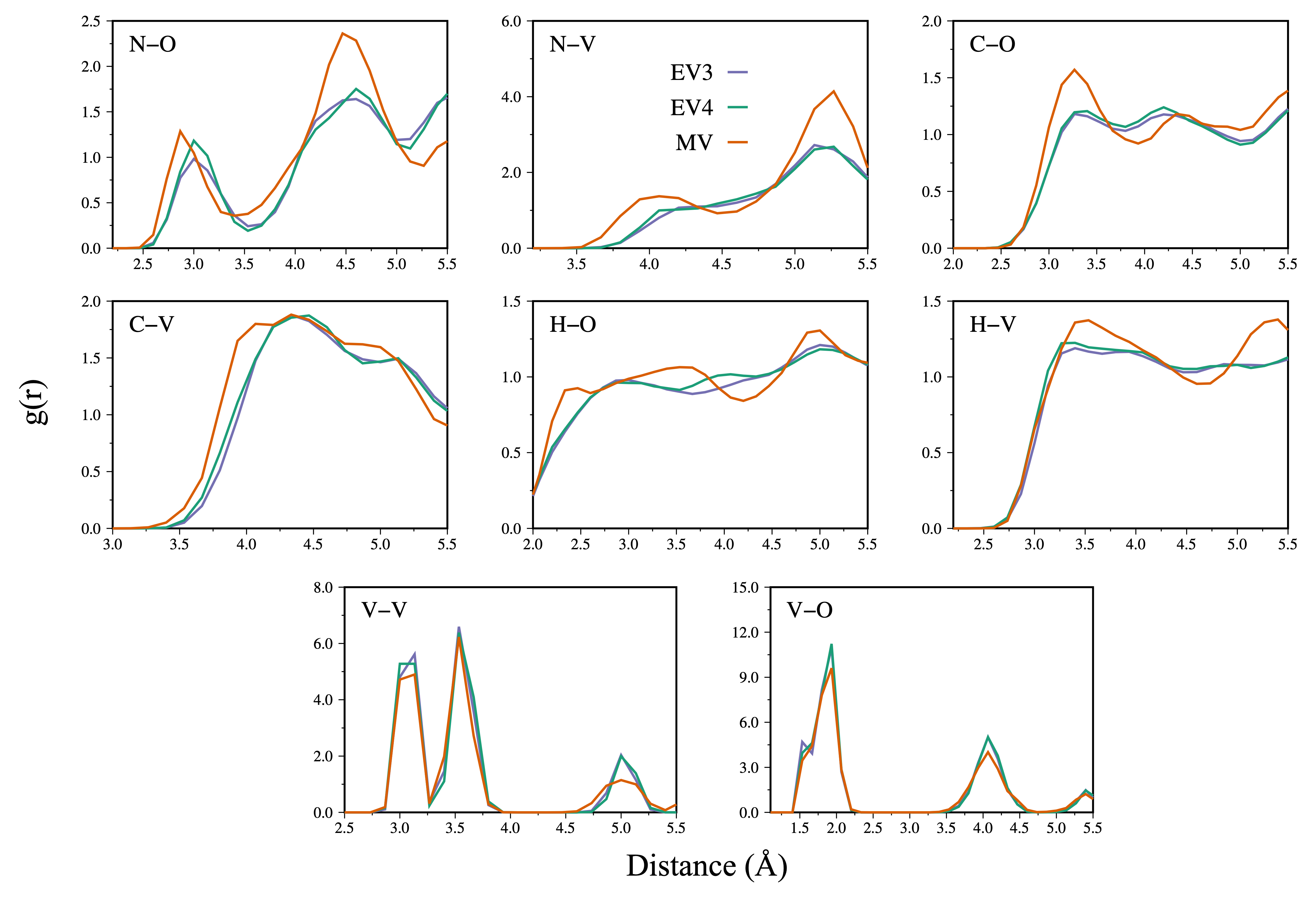}
    \caption{Calculated RDFs of all studied systems using AIMD simulations at 298.15 K in the NVT ensemble.}
\addcontentsline{toc}{subsection}{Figure \ref{fig:rdf_all}. Calculated RDFs of all studied systems using AIMD simulations at 298.15 K in the NVT ensemble.}
    \label{fig:rdf_all}
\end{figure*}

\clearpage
\begin{figure*}[!h]
    \centering
    \includegraphics[width=0.99\linewidth]{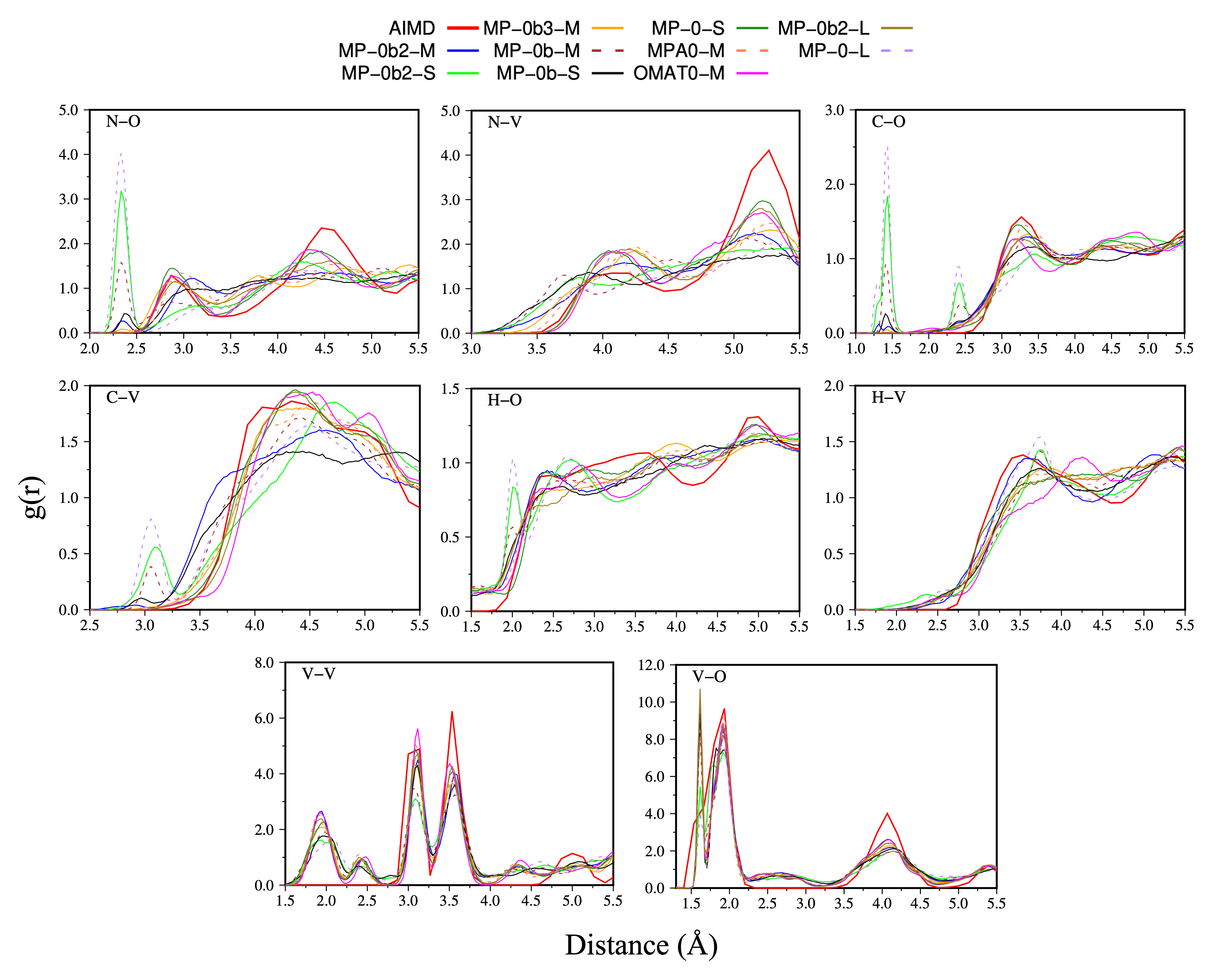}
    \caption{Calculated RDFs using different MACE models compared to the reference AIMD data in the NVT ensemble at 300 K for the representative MV system.}
\addcontentsline{toc}{subsection}{Figure \ref{fig:mace_bench}. Calculated RDFs using different MACE models compared to AIMD for MV.}
    \label{fig:mace_bench}
\end{figure*}

\clearpage
\begin{figure*}[!h]
    \centering
    \includegraphics[width=0.99\linewidth]{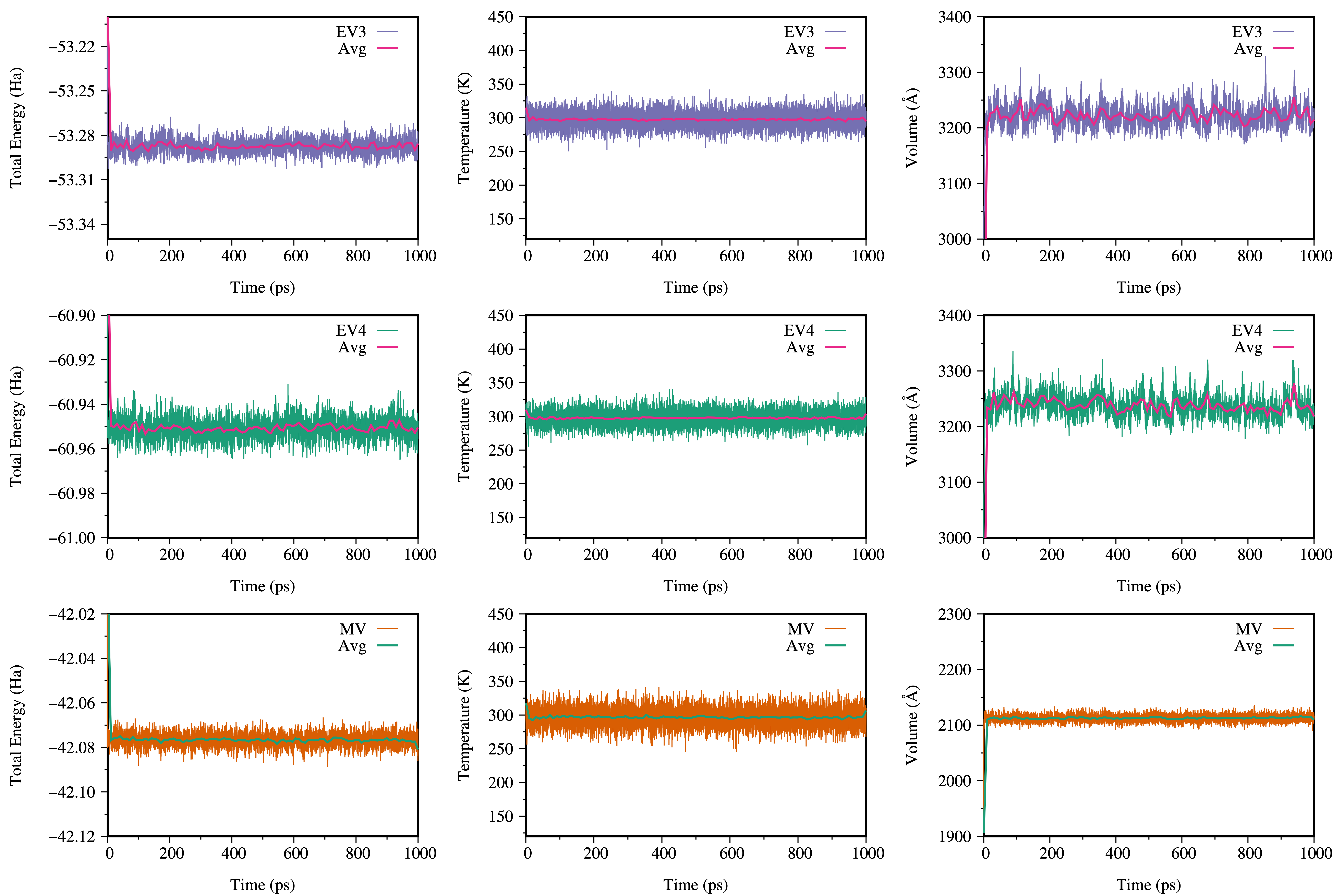}
    \caption{Convergence of the energy (in Hartree), temperature (in kelvin), and volume (in \AA$^3$) for the EV3, EV4, and MV systems at 300 K and 1 atm pressure in the NPT ensemble as obtained from MACE-MP-0-Small simulations. The moving 50 points average is shown in red.}
\addcontentsline{toc}{subsection}{Figure \ref{fig:mace-conv}. MACE convergence plots.}
    \label{fig:mace-conv}
\end{figure*}

\clearpage
\begin{figure*}[!h]
    \centering
    \includegraphics[width=0.99\linewidth]{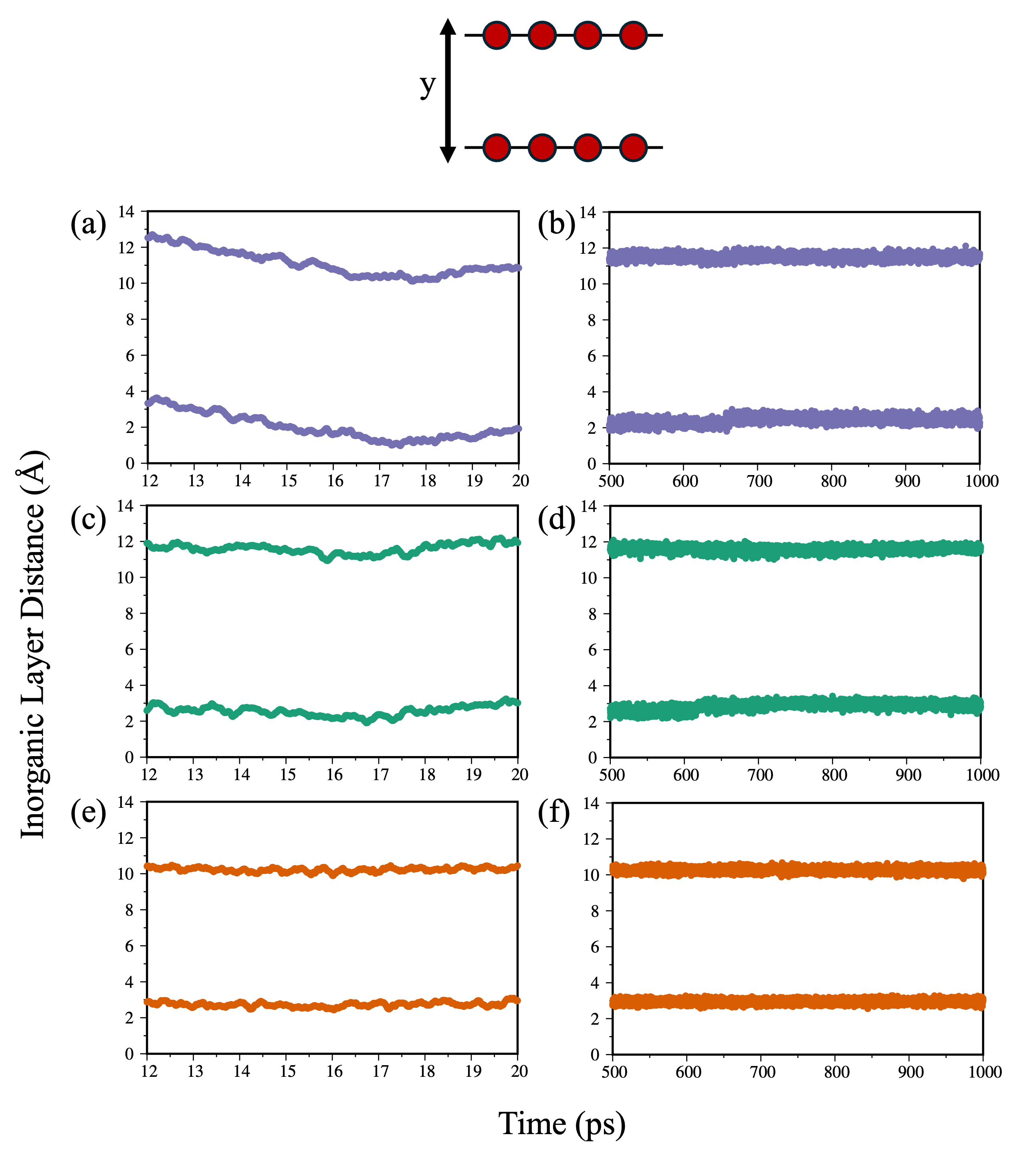}
    \caption{Validating the performance of the MACE-MP-O-Small model in calculating HB inorganic flexibility against the reference AIMD simulations. AIMD data (a,c,e) for EV3 (top), EV4 (middle), and MV (bottom) are compared to the corresponding MACE results (b,d,f).}
\addcontentsline{toc}{subsection}{Figure \ref{fig:flex}. Validating the performance of the selected MACE model in calculating HB flexibility against reference AIMD.}
    \label{fig:flex}
\end{figure*}

\clearpage
\begin{figure*}[!h]
    \centering
    \includegraphics[width=0.6\linewidth]{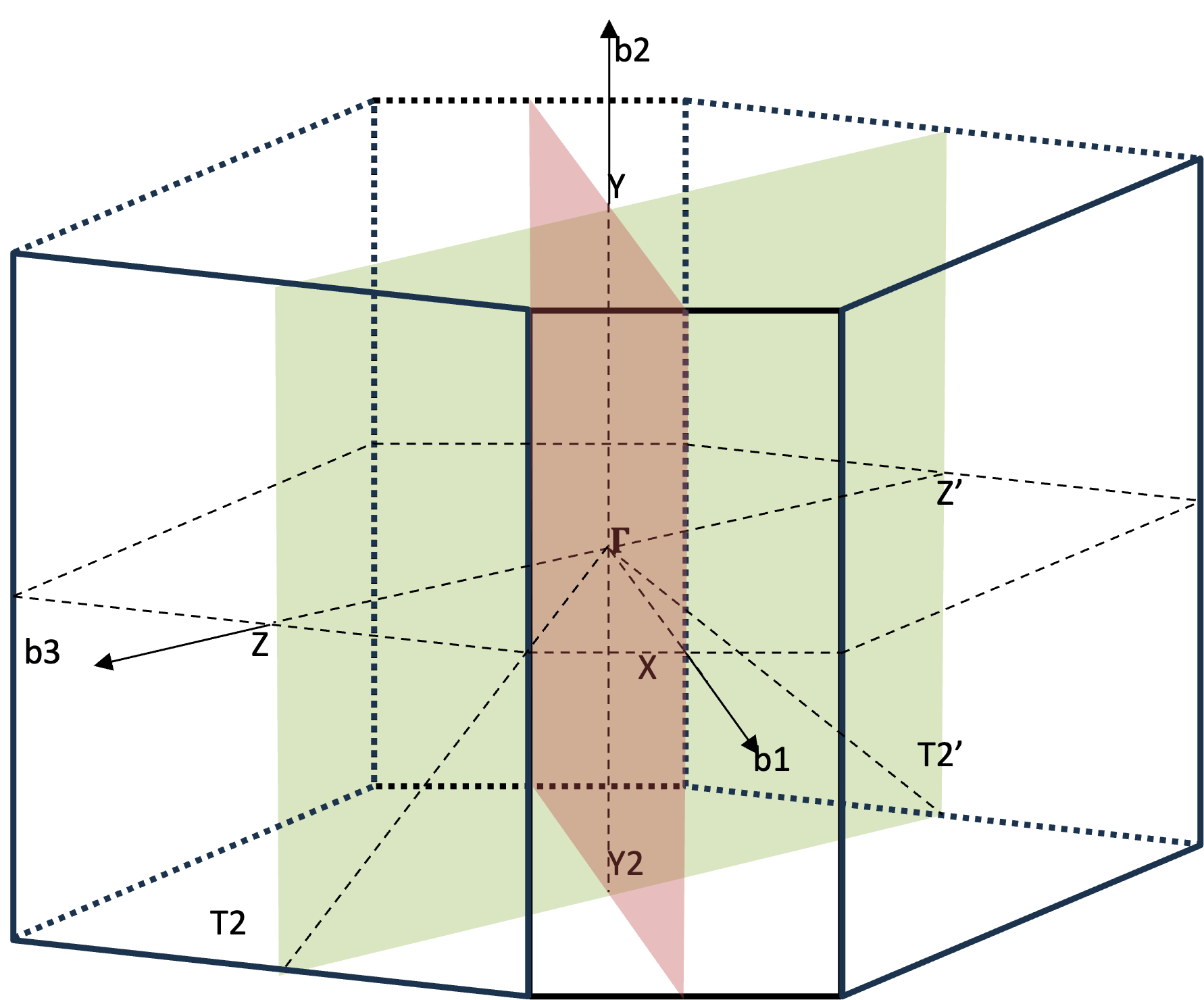}
    \caption{Brillouin zone and k-point paths used in our band structure calculations. The green plane depicts the in-plane, while the red plane corresponds to the out-of-plane pathways for all systems.}
\addcontentsline{toc}{subsection}{Figure \ref{fig:k-path}. Brillouin zone and k-path for the band structure calculations.}
    \label{fig:k-path}
\end{figure*}

\clearpage
\begin{figure*}[!h]
    \centering
    \includegraphics[width=0.79\linewidth]{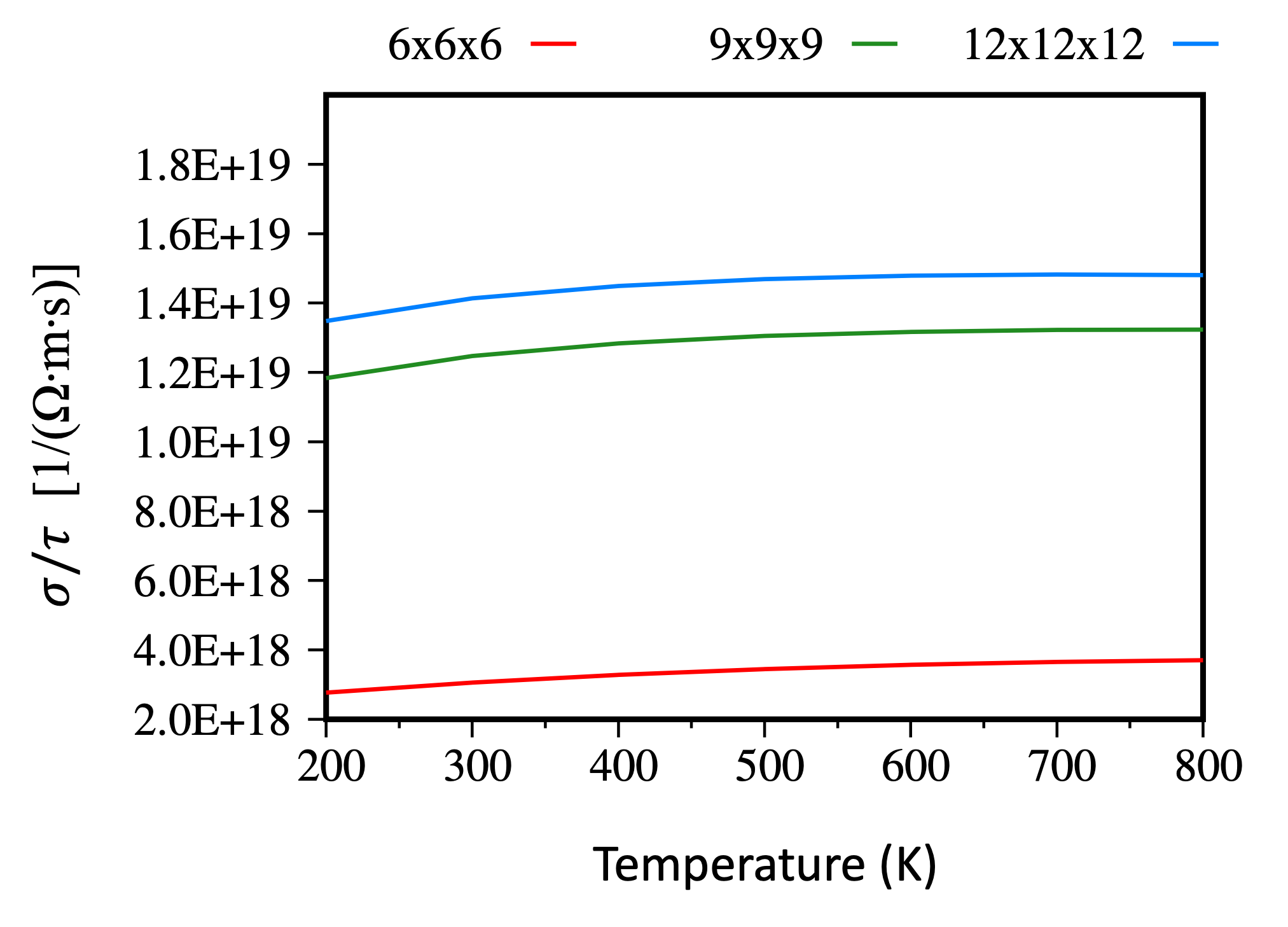}
    \caption{Convergence of the calculated electrical conductivity of the representative MV system using VASP and BotzTraP2.}
\addcontentsline{toc}{subsection}{Figure \ref{fig:conductivity_convergence}. Benchmark of electrical conductivity convergence for MV.}
    \label{fig:conductivity_convergence}
\end{figure*}

\clearpage
\bibliography{bib} 
\addcontentsline{toc}{subsection}{References.}